\documentclass[useAMS,usenatbib]{mn2e}

\usepackage{graphicx}

\def\gsim{ \lower .75ex \hbox{$\sim$} \llap{\raise .27ex \hbox{$>$}} } 
\def\lsim{ \lower .75ex \hbox{$\sim$} \llap{\raise .27ex \hbox{$<$}} }

\title[Multi-phase cooling flows within galactic haloes]{ 
Cooling flows within galactic haloes: the kinematics and properties of  
infalling multi-phase gas} 
\author[Kaufmann et al.] 
{Tobias Kaufmann $^1$ \thanks{E-mail: tkaufmann@physik.unizh.ch},  
 Lucio Mayer $^1$, James Wadsley $^2$, Joachim Stadel $^1$\newauthor  
{and Ben Moore $^1$  } 
\\$^1$ Institute for Theoretical Physics, University of Z\"urich, CH-8057 Z\"urich, Switzerland 
\\$^2$ Department of Physics \& Astronomy, McMaster University, 1280 Main St. 
West, Hamilton ON L8S 4M1 Canada} 
 
\begin{document} 
 
\date{Accepted year  Month day. Received year Month day; in original form year Month day} 
 
\pagerange{\pageref{firstpage}--\pageref{lastpage}} \pubyear{}  
 
\maketitle 
 
\label{firstpage}

\begin{abstract} 
We study the formation of disks via the cooling flow of gas within  
galactic haloes using smoothed particle hydrodynamics 
simulations. These simulations resolve mass scales of a few thousand solar masses in the gas component for the first time. Thermal instabilities 
result in the formation of numerous warm clouds that are 
pressure confined by the hot ambient halo gas.  
The clouds fall slowly onto the disk through non-spherical accretion 
from material flowing preferentially down the angular momentum axis. 
The rotational velocity of the infalling cold gas decreases as a function of  
height above the disk, closely resembling that of the extra-planar gas  
recently observed around the spiral galaxy NGC 891. 
 
\end{abstract}

 
\section{Introduction} 
 
Galaxy formation is a complex process involving the simultaneous action of 
many physical mechanisms. Even though we have a well defined cosmological 
framework, $\Lambda$CDM,
within which to study galaxy formation, numerical simulations  
have achieved limited success as far as reproducing the structural properties 
of observed galaxies. Such simulations  
give rise to disks that are  
smaller and denser than their observed counterparts 
(e.g. Navarro \& Benz 1991, Katz et al. 1992,  Navarro \& Steinmetz 2000). A further problem is the difficulty in 
producing disk dominated systems, even when haloes with quiet merger histories 
are selected. 
These problems may 
be due to the lack of a correct treatment of the physics of the multiphase  
interstellar medium, in particular of the balance between radiative cooling 
and heating from various forms of feedback arising from star formation 
(White \& Frenk 1991,  Navarro \& Benz 1991, Katz et al. 1992, Robertson et al. 2004). However, limitations in the numerical models, 
for example the coarse mass and force resolution, are perhaps another 
major cause of the problem (Governato et al. 2004).  
 
In this paper we use new N-Body+SPH simulations where we achieve a resolution in the 
gas component of $730$ M$_{\odot}$, in order 
to follow the formation of a galactic disk via the cooling of gas within 
equilibrium dark matter halos. Cosmological simulations 
suggest that the large disks of spiral galaxies form mainly from 
the smooth accretion of gas after the last major merger  
(Sommer-Larsen, G\"otz \& Portinari 2003, Abadi et al. 2003, Governato et al. 2004). 
Therefore, although our simulations are not within the full cosmological  
framework, they are designed to follow the quiet gas accretion phase during the main 
epoch of disk formation. This allows us to resolve the cooling flow of gas 
at a very high resolution, which complements studies of 
disk formation within cosmological simulations.  
 
The hierarchical formation of massive dark matter  
haloes, above a characteristic circular velocity of  
$120$ km/s is expected to efficiently shock heat the gas to their virial 
temperatures throughout the virial region  
(Dekel \& Birnboim 2004, McCarthy et al. 2003). 
Angular momentum of the gas and dark matter  
is generated early by tidal torques, allowing it to cool and form a  
rotationally supported disk (Crampin \& Hoyle 1964). 
The process of gas cooling from the halo into the disk is difficult to study 
observationally and little evidence for gas accretion is observed in galaxies or cluster 
mass haloes. Cooling flow clusters are so named because of the observed  
decrease in the 
central temperature of the gas. In galactic haloes the gas temperature can be  
lower than $10^6$ K and thus difficult to observe by 
current X-ray telescopes (Benson et al. 2000), but there is evidence for a hot ionised corona  
surrounding the Milky Way. 
 
Recent measurements of OVI and OVII absorption in the UV and X-ray 
part of the spectrum (Sembach et al. 2003) 
lend support to the idea that a lumpy gaseous Galactic 
halo exists and that such emission comes from the interface between warm clouds 
and the hotter diffuse medium. Maller \& 
Bullock (2004) suggested that the gas supply to galaxies is 
mostly in the form of discrete warm clouds.  
Small density and temperature fluctuations are enhanced by the cooling 
process resulting in a runaway instability and the formation of  
a fragmented distribution  
of cooled material, in the form of warm $(\sim10^{4}$K$)$ clouds, 
pressure-supported within the hot gaseous background. This model may explain 
the properties of high velocity clouds (HVCs) around the Milky Way (Blitz et al. 1999), for 
example their radial velocity distribution and angular sizes.  
 
Recently, Fraternali et al. (2005) measured the rotation 
curve of extra-planar  
neutral gas around the large spiral galaxy NGC 891. This gas could be accreting  
material or gas falling back to the disk via a galactic fountain from star-formation. 
The thermal instability mentioned above results in  
cloud masses that are predicted to be $\sim 10^6 $ M$_{\odot}$ which is an order of magnitude below the minimum 
mass resolved in the SPH calculation in current state-of-the art galaxy  
formation simulations (e.g. Governato et al. 2004). 
In this paper we aim to have sufficient resolution to 
study the formation of a two-phase medium which will allow us to  
compare the kinematics of the gas with the observations  
of Fraternali et al. 
 
The plan of this paper is as follows: In Section 2 we present 
the modelling of our halos, the treatment of the cooling 
and the numerical techniques. 
In Section 3 we compare the kinematics of the gas and the disk 
with the observational data from Fraternali et al. The properties 
of the cool clouds are studied in Section 4, including survival 
times, mass function and numerical convergence studies. We conclude 
and summarise in section 5. 
 
\section{Initial Conditions and Methods} 
 
\begin{figure} 
\includegraphics[
  scale=0.68] 
{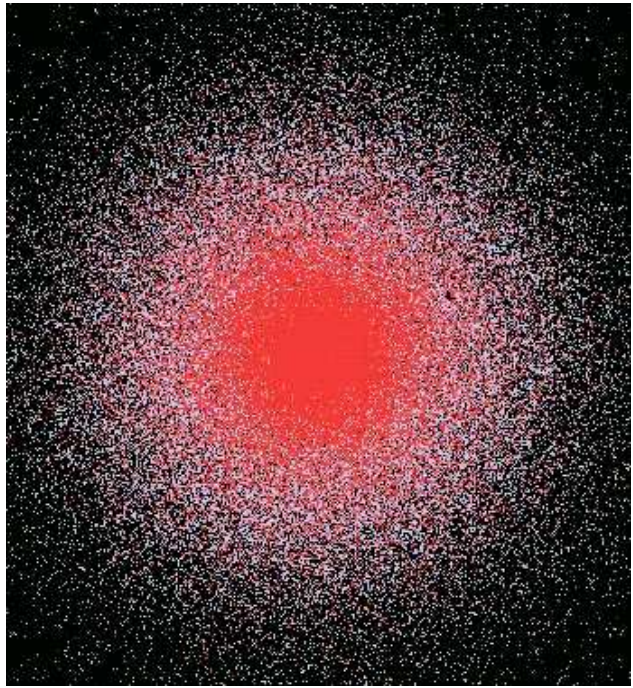}  \includegraphics[
scale=0.4]{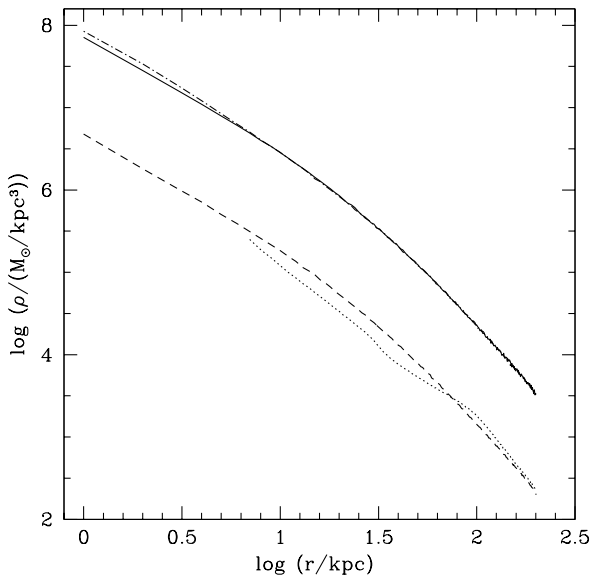}

\caption{\label{ics}The upper panel shows the initial particle distribution (dark matter in white, gas in red,  the box has a side length of $1200$ kpc) and the lower panel the radial density profiles of gas and dark matter: the density of the dark matter initially is shown with a solid line whereas the dotted-dashed line shows it after  $0.5$ Gyr. The dashed line shows the initial gas density and the dotted line the density of the hot gas outside the disk region.}
\end{figure}

We set up a spherical equilibrium halo with an NFW profile (Navarro, Frenk \& White 1996) 
and structural parameters consistent with predictions of the   
standard $\Lambda$CDM model (Kazantzidis, Magorrian \& Moore 2004). 
We include a fraction of the total halo mass, $f_b$, as a hot 
baryonic component with the same radial distribution and a temperature 
profile such that the gas is initially in hydrostatic equilibrium for an adiabatic equation of state (EOS) for the gas. Of course this kind of setup can not model galaxy formation in general (no substructure, no counter rotation etc.) but should reproduce the late phase accretion after the last major merger.   To illustrate the initial conditions we plot the initial particle distribution and the density profiles in Figure \ref{ics}. The shape for the initial density of gas and dark matter is the same whereas only the normalisation is different. The curves after $0.5$ Gyr show that the hot halo evolves out of equilibrium  when using cooling, the density of the dark matter after  $0.5$ Gyr is enhanced in the centre due to adiabatic contraction.

The dark matter haloes are constructed using a multi-mass technique 
such that the innermost region is very well resolved. 
For the M33 ``standard'' model (see below) we use $1,400,000$ particles distributed in the inner sphere of radius $20$ kpc,  
then $600,000$ particles in the next shell to $100$ kpc and the outer 
halo is resolved with $200,000$ particles. This minimises  
spurious heating of the gas particles due to collisions with dark matter 
particles (Steinmetz \& White 1997) and enables us to resolve 
the dark matter cusp to $\sim100$ parsecs. The reliability of these 
initial conditions are tested and reported in Zemp et al. (2006, in preparation).

The gas has a specific angular momentum distribution and spin 
parameter consistent with values found for dark matter haloes 
within cosmological N-body simulations. The initial specific
angular momentum profile is a power law following 
$j\propto r^{1.0}$, similar to the value   
$j(r)\propto r^{\alpha},\alpha=1.1\pm0.3$ found by Bullock et al. (2001). 
We also ran simulations with different angular momentum distributions to 
confirm that this does not affect the conclusions in this paper. 
The spin parameter is $\lambda=\frac{j_{gas}\left|E\right|^{0.5}}{GM^{1.5}}$, 
where $j_{gas}$ is the average specific angular momentum of the gas, $E$ is 
total energy of the halo and $G$ is the gravitational constant. This 
definition matches the one commonly used under the assumption that 
there is no angular momentum transport between the spherical dark 
matter halo and the gas. We carried out additional simulations where  
we set up angular momentum profiles by 
merging two equilibrium spherical haloes with different impact  
parameters. Both sets of simulations gave similar results. 
The detailed description of the runs are presented in Kaufmann et al. (2006). 
 
We constructed dark plus gaseous halo models with parameters that are expected 
to produce disks similar to those of the Milky Way (MW model) and NGC 598 (M33 model). 
For the MW model the virial velocity $v_{vir}=140$ km/s, virial radius  
$r_{vir}=200$ kpc, virial mass $M_{vir}=9.14\times10^{11}$ M$_{\odot}$, 
halo concentration $c=8$, spin parameter  
$\lambda=0.038$ and the baryonic fraction is $f_{b}=9\%$ and we resolve the gas halo with particles of  $\sim2\times10^{5}$ M$_{\odot}.$ 
 For the M33 model the parameters are: $M_{vir}=5\cdot10^{11}$ M$_{\odot}$, $r_{vir}=167$ kpc, $v_{vir}=115$ km/s, $c=6.2$, $f_{b}=6\%$ and spin parameter $\lambda=0.105$.    In the standard M33 model the hot gaseous halo is resolved with $2\times 10^6$ particles of equal mass $\sim2\times10^4$ M$_{\odot}$. To test further the effects of resolution, we performed simulations in which we use eight times as many SPH particles within the inner 30 kpc, the ``refined\_8'' M33 simulation. We split each SPH particle into eight ``child'' particles  (Bromm 2000, Kitsionas 2000, Escala et al. 2004). The new particles are randomly distributed according to the SPH smoothing kernel within a volume of size $\sim h_p^3$, 
where $h_p$ is the smoothing length of the parent particle. The velocities 
of the child particles are equal to those of their parent particle and so is  
their temperature, while each child particle is assigned a mass equal to  
$1/N_{split}$ the mass of the parent particle. The new particles have a mass of $\sim 3000 $ M$_{\odot}$.  We used then the same method to create the  ``refined\_32'' M33 simulation, where each SPH particle within the inner 30 kpc has been split into $32$ particles having masses of $\sim 730 $ M$_{\odot}$. An overview to the M33 simulations can be found in Table \ref{cap:tab mass 2}.

We use the high performance parallel Tree+SPH code GASOLINE (Wadsley, Stadel \& Quinn 2004), which is an extension of the pure N-Body gravity code 
PKDGRAV developed by Stadel (2001). It uses an artificial viscosity which is the shear 
reduced version (Balsara 1995) of the standard Monaghan (1992) artificial 
viscosity.  GASOLINE uses a spline  
kernel with compact support for the softening of the gravitational 
and SPH quantities. The energy equation is solved using the asymmetric formulation,
\begin{eqnarray}
\frac{d u_i}{dt}& = & \frac{P_i}{\rho_i^2} \sum_{j=1}^{n}{m_j}
{\bf v}_{ij} \cdot \nabla_i W_{ij},
\end{eqnarray}
where $u_i$, $P_i$, $\rho_i$, $m_i$, ${\bf v}_{ij}$ and  $W_{ij}$ are the internal energy, pressure,  density, mass, velocity  and a kernel function, respectively. This formulation is shown to yield very similar results compared to the entropy conserving formulation but conserves energy better (Wadsley, Stadel \& Quinn 2004).
 In a test problem with an expanding sphere and no gravity, the entropy   conservation for each particle as measured by the temperature error has  an RMS error of $3\%$ for a factor of $10$ change in density; however, 
this is a very conservative estimate since in situations of astrophysical interest
entropy conservation should be measured  against the other contributors to changes in entropy, namely 
shocks  and cooling, and these are expected to be dominant, especially in the case of a gravitational
collapse. In all the runs we included Compton and radiative cooling 
using a standard cooling function for a fully ionised, primordial 
gas of helium and hydrogen. The radiative losses per step are limited to no more than $25 \%$  per 
dynamical  step and the cooling can adapt the time steps to suit its needs. The density changes are quite slow -- occurring on the dynamical time  scale and typically less than 1-2 percent per step.  Therefore the assumption of a constant density in one time step contributes only  a small error to the cooling rates, especially compared to missing coolants such as metals.
Because of the lack of molecular 
cooling and metals the efficiency of our cooling function drops sharply 
below $10,000$ K. Moreover we used an artificial lower 
limit for the gas temperature ($15,000$ K or $30,000$ K), higher than 
the cut-off in the cooling function, to 
crudely model the effect of the UV background and stellar feedback 
(see e.g. Barnes 2002). This temperature floor is also required to avoid 
the onset of a violent gravitational instability and fragmentation  
of the gaseous layer after the disk begins to assemble.

For the identification of the clouds we use a friends of friends algorithm (FOF), using $32$ particles (equal to the number of particles in the smoothing kernel) as the threshold. The linking length was chosen to be of order $0.3$ the mean particle separation in the standard run, it  was then lowered for higher resolution (according to the smaller mean particle separation) and we checked in the standard simulation that the number of clouds found with the FOF does depend only weakly on the exact choice of the linking length.

\section{The assembly of the disk} 
 
\begin{figure} 
\includegraphics[%
  scale=0.82] 
{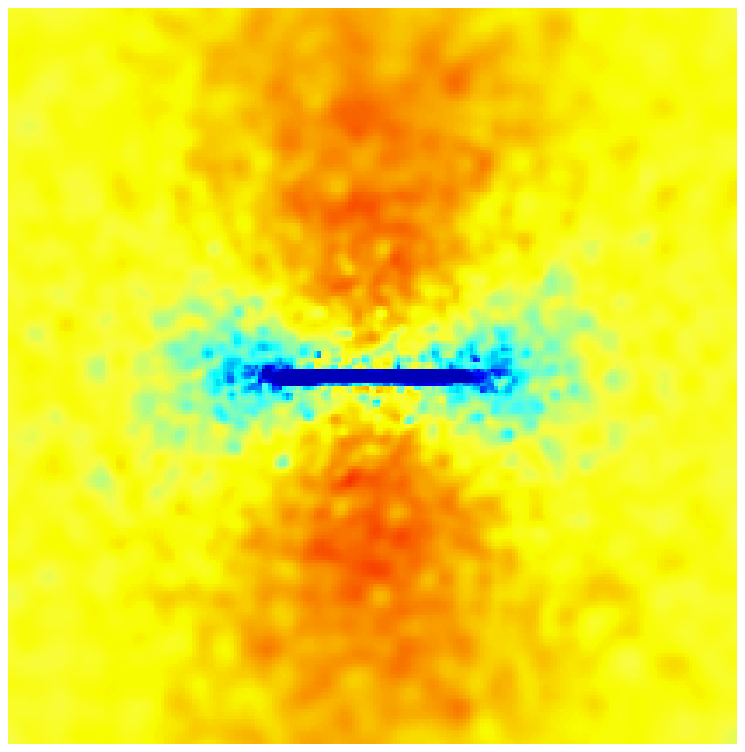}  \includegraphics[%
scale=0.511]{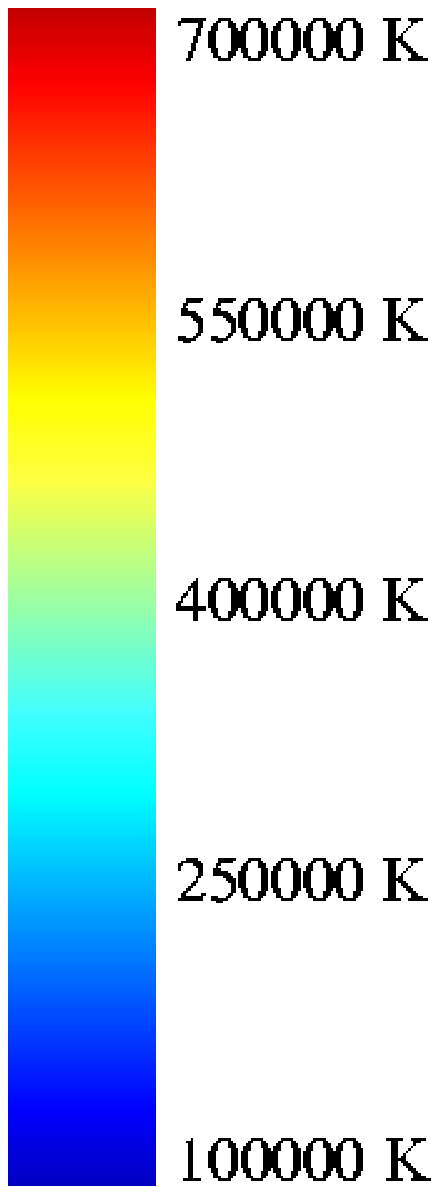} 
\includegraphics[%
scale=0.533]{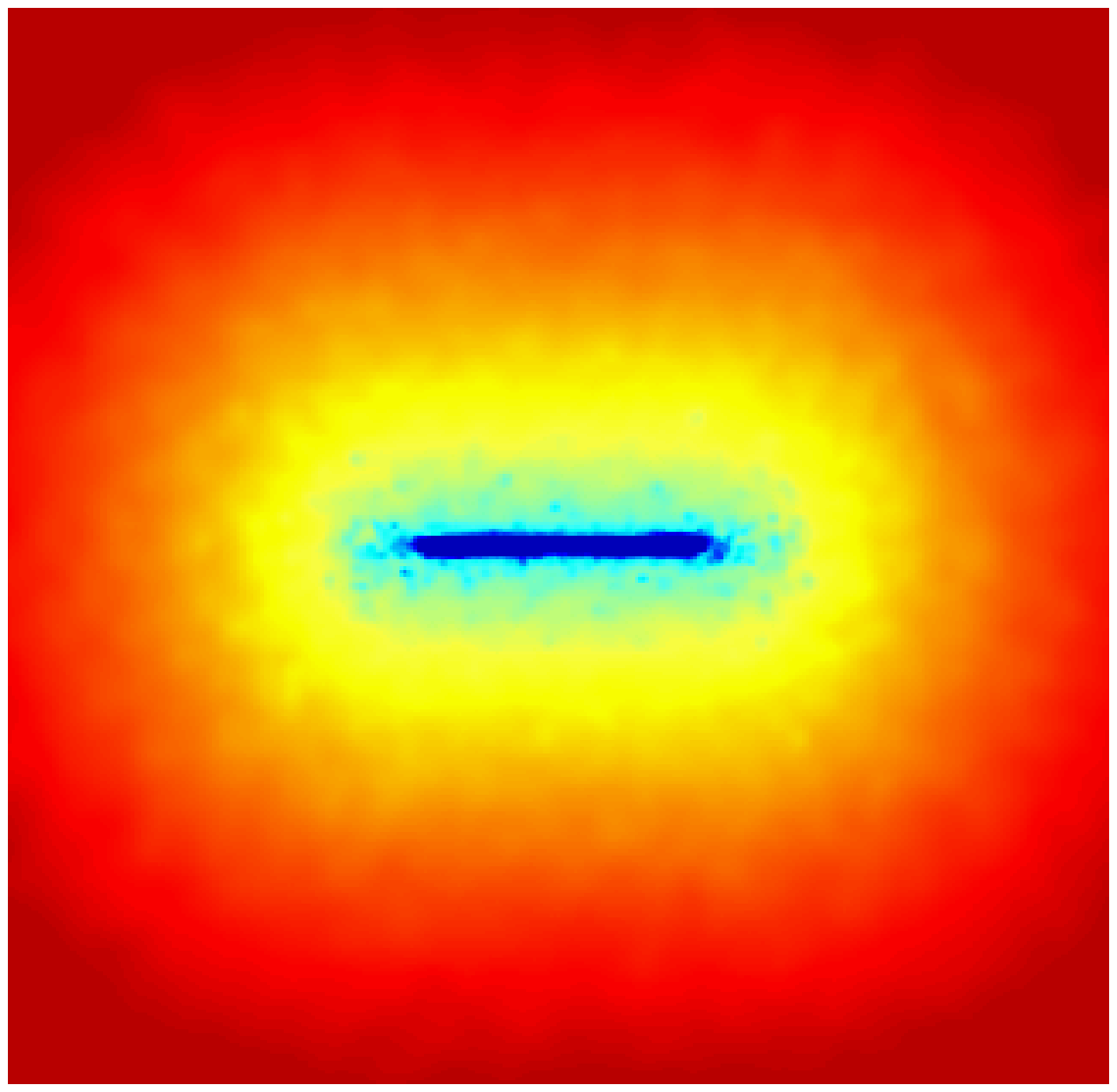} \includegraphics[%
scale=0.358]{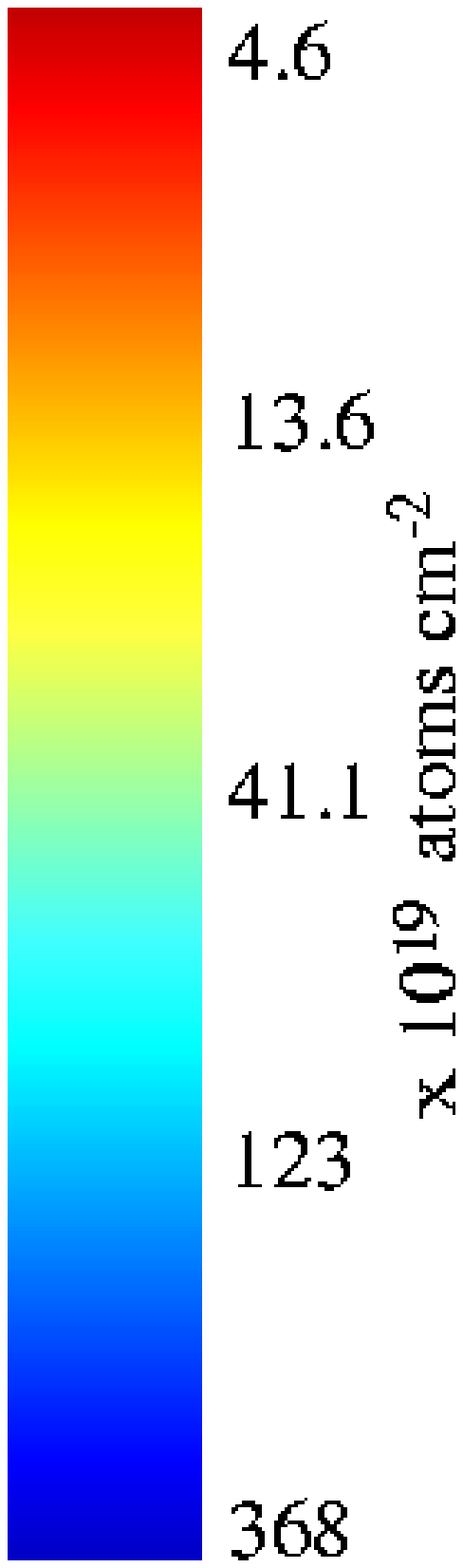}\ 
 
\caption{\label{cap:firstcoll}The two panels show colour maps of projected  
temperature (upper map) and density (lower panel) after 2.1 Gyr of the
standard M33 model (see text). 
The box has a side length of $60$ kpc. The hotter and  
less dense gas above and below the disk is visible, see text for details.} 
\end{figure}

Due to cooling the hot gas halo loses its hydrostatic equilibrium configuration quite quickly (see Figure \ref{press-sup}). The inner disk rapidly forms from cooling 
gas from a nearly spherical region close to the  
centre (within $\sim 10$ kpc) of the halo. 
The accretion rate of cold gas slows down with time and 
the outer disk region forms from material which flows preferentially down 
the angular momentum axis. This leaves a ``cylindrical'' region of  
hotter and less dense gas above and below the disk plane as is evident in   
Figure \ref{cap:firstcoll}. This non-spherical inflow of gas from the halo 
to the disk is not currently considered in semi-analytic models for disk formation. 
 
As expected, the higher angular momentum M33 model produces a more extended disk 
than the MW model. Furthermore the morphology of the disks within the two 
models is very different. The higher baryon fraction and lower spin parameter 
of the MW model gives rise to a disk that dominates the mass distribution 
in the inner region. The MW model forms a bar in the very early stages and has the 
appearance of an Sb galaxy whereas the M33 model resembles a galaxy of type  
Sc-Sd, with a smoother more flocculent spiral pattern. 
Even though the M33 model does not undergo bar formation, the cooling gas forms a dense central nucleus which has a size of the order our length 
resolution of a few hundred parsecs. The global morphological evolution 
of both models depends on the mass and force resolution  
and is discussed  in Kaufmann et al. (2006). 
Here we present the analysis of the kinematics of the cooling gas. 
 
\subsection{Gas infall and extra planar gas} 
 
Fraternali et al. (2005) observed the rotation velocity of neutral gas as 
a function 
of scale height above the disk for the galaxy NGC 891 which has a luminosity 
and rotational speed similar to the Milky Way. The found extra planar 
neutral gas up to distances of 15 kpc above the plane and found a 
velocity gradient of about $-15$ km s$^{-1}$kpc$^{-1}$ in the vertical direction.  
The gas closest to the disk, at
$z<1.3$ kpc, appears to corotate with it 
but they argue that due to the limited 
angular resolution this may be the effect of beam-smearing. They considered 
a galactic fountain model in order to explain these observations, but that model failed to reproduce the amount of lagging. They argue that gas accretion may be important as well in these situations. 
 
\begin{figure} 
\includegraphics[%
  scale=0.45] 
{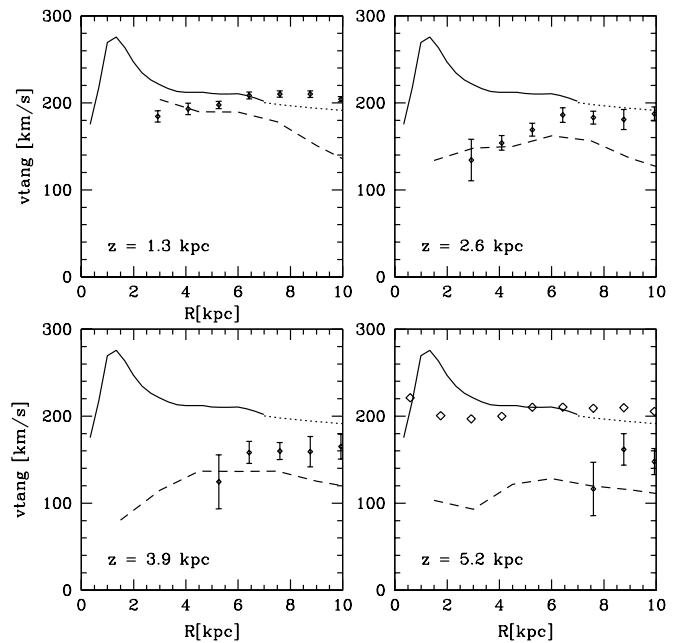} 
 
\caption{\label{cap:rotcurve}The solid lines show the 
physical rotational velocity of the gas in the disk plane of the MW model. 
The dashed lines in the different panels show the rotational speed at  
various distances above the galactic plane. The vertical 
velocity gradients in the model  
are $16,21,19$ and $17$ km s$^{-1}$kpc$^{-1}$ respectively. 
The open symbols in the last panel  
show the observed rotational speed of NGC 891 in  
the disk plane and the points with error bars in  
the panels show the observations 
above the plane of the galaxy. The rotation  
curve in the disk of NGC 891 is scaled such  
that the flat part matches the  rotation  
curve in the disk plane of our simulation; the rest  
of the data and also height $z$ above the disk  
are then scaled with the same factor. 
} 
\end{figure}

Here we present results from our MW simulation where 
we show that this velocity gradient arises naturally in the accretion 
phase of the gas as it falls onto the  
galactic disk.  
Figure \ref{cap:rotcurve} shows the rotational velocity of the 
gas as a function of height above the disk in our MW simulation. We plot the actual  circular rotational speed of the gas averaged over the annulus at the given radii. (Note that  
this is slightly different to 
the observations which measure  
the projected rotational speed at a given radius.) 
The central velocity curve 
has a bump due to the bar formation and the excess of low 
angular momentum gas in the inner halo. The decrease in rotational velocity above 
the plane matches the observations quite well. This suggests that the 
observations 
of Fraternali et al. support the idea that disks form via the cooling flow 
of gas which is rotating faster closer to the disk by conservation of angular momentum. 
 
\begin{figure} 
\includegraphics[%
  scale=0.45]{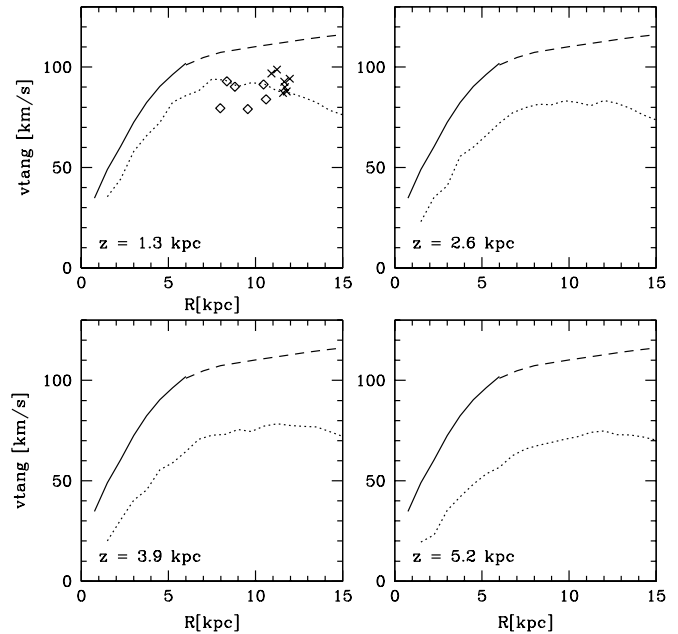}

\caption{The solid lines show the physical rotational velocity of the gas in the disk plane of the M33 model  after 2 Gyr (dashed lines show circular velocities). The dotted lines in the different panels show the rotational speed at 
various distances above the galactic plane of all the gas in the respective plane.  The open symbols show the tangential velocities of the cold clouds near the disk, which are between $z=1.3$ kpc and  $2.4$ kpc away from the galactic plane, and the crosses mark clouds which are at distances less than  $1.3$ kpc. The calculated velocity gradients (from all the gas) are
$8,8,6$ and $6$ km s$^{-1}$kpc$^{-1}$ respectively, whereas we get $6$ km s$^{-1}$kpc$^{-1}$ for the velocity gradient derived from the clouds. \label{cap:Rotation-curves-M33}} 
\end{figure}

\begin{figure} 
\includegraphics[%
  scale=0.4]{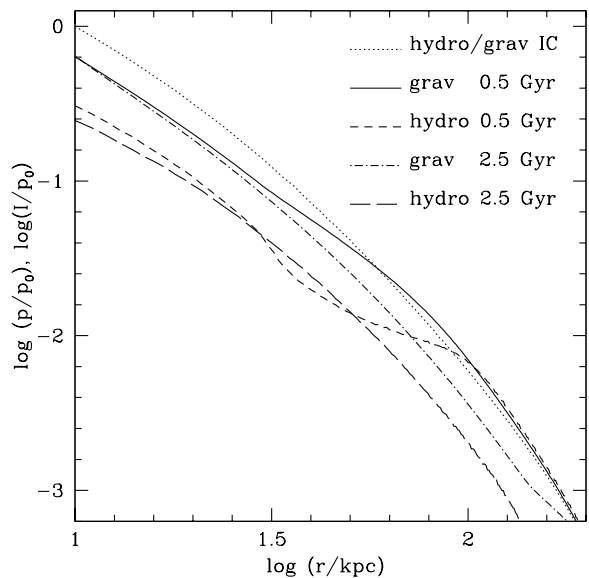}

\caption{The time-evolution of the pressure in the hot gas halo and the gravitational attraction $(I\equiv\int_{R}^{\infty} G \rho_g(r)\frac{GM(r)}{r^2} \,dr)$   is plotted versus radius (standard M33 run). Initially, gravity and pressure were balancing each other, at later times the gas halo has lost pressure support due to cooling.  \label{press-sup}} 
\end{figure}

In Figure \ref {cap:rotcurve} we  plot the rotational speed of all the gas as a function 
of position above the disk plane, not just gas in the neutral phase such as measured 
by Fraternali et al.  However, in  a recent observation of NGC 891 by Heald et al. (2005) it has been found that the velocity gradients of the hot gas agree with the results for the neutral gas. As we then increase the resolution we find 
that the hot halo is thermally unstable and forms dense  
gas clouds which then accrete onto the disk. These  clouds may cool 
down and become partially neutral (see section 4.2.3). 
We only achieve the resolution 
necessary to observe this two phase medium in our M33 model (see Figure \ref{cap:Rotation-curves-M33} for the corresponding velocity plot) but
such clouds would occur also in the MW model if we had sufficient resolution. 
Unfortunately, due to the higher baryonic density in the centre of the 
MW model this simulation is much more demanding computationally since 
particles require shorter time steps. The cool clouds in the M33 rotate at comparable speed as all the gas, which justifies the measurement of the kinematics within all the gas.  The determination of the velocity gradient from the cold clouds alone was only possible in the inner region,  because there were too few clouds at higher $z$ in the standard M33 model, but we get a comparable number as if we take all the gas in the plane (Figure \ref{cap:Rotation-curves-M33}).  Figure \ref{press-sup} shows that the hot gas halo can not preserve pressure-support for an EOS with cooling. The infall of the hot gas is therefore expected.

\begin{figure} 
\includegraphics[%
  scale=0.4]{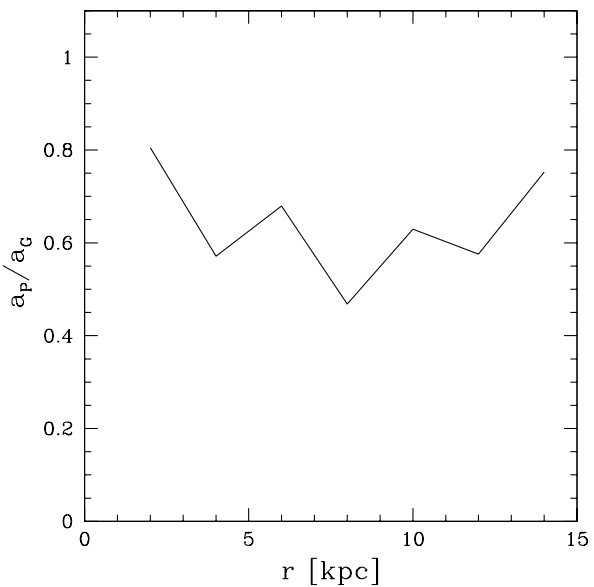}

\caption{The ratio between hydrodynamical and gravitational acceleration in the $z$-direction (perpendicular to the disk) is plotted versus radius. For this measurement gas particles in a box were chosen, the box is  $2$ kpc above the disk centre and has a height of $1$ kpc; the model is $2$ Gyr old. In average the hydrodynamical acceleration is $\approx 60 \%$ of the  gravitational acceleration \label{press-grav}} 
\end{figure}

The hydrodynamical forces due to  diffuse hot corona on the infalling gas are substantial: the infall velocity in the M33 galaxy of the gas above the disk is approximately $10$ km/s. This is much smaller than the pure free-fall 
velocity towards the disk, $\sim 70$ km/s,  This radial infall velocity  
agrees well with that observed by Fraternali et al. (2002) of   
approximately $15$ km/s for NGC 2403, a galaxy similar to M33. In Figure \ref{press-grav} we show the ratio between pressure (hydrodynamical) and gravitational accelerations perpendicular to the disk for some gas particles $2$ kpc above the disk. The pressure forces can equilibrate the gravitational forces in  average at the  $\approx 60 \%$ level, which results in this lowered infall velocity. The result for particles which are e.g. $20$ kpc above the disk is similar.

\section{Disk formation via accretion of a clumpy medium} 
 
Maller and Bullock (2004, hereafter MB04) proposed that disks form via the accretion of warm 
clouds that develop from thermal instabilities in the hot gaseous halo (see also Mo \& Miralda-Escude 1996).
This instability arises as 
small density fluctuations cool faster than most of the surrounding gas; 
cooling amplifies these density fluctuations further and the instability 
rapidly grows. According to MB04 this clumpy formation of disk galaxies would  
at the same time explain the existence of high velocity clouds 
and give rise to a characteristic upper 
limit on the masses of galaxies consistent with observations. 
Typical parameters for 
the clouds in a Milky Way-sized halo today would be a size of  
$\sim 1$  kpc and a characteristic  
cloud mass of $\sim5\times 10^{6}$ M$_{\odot}$.  
Several physical mechanisms, e.g. the Kelvin-Helmholtz instability and
conduction, impose a lower mass limit on clouds that can survive (see below). 
These clouds have not been found in cosmological simulations of galaxy 
formation since even the highest resolution studies have a single SPH particle mass equal to the expected characteristic cloud mass (Governato et al. 2004; Sommer-Larsen et al. 2003). 
 
\subsection{Formation of pressure confined clouds} 
 
\begin{figure} 
\includegraphics[%
  scale=0.80]{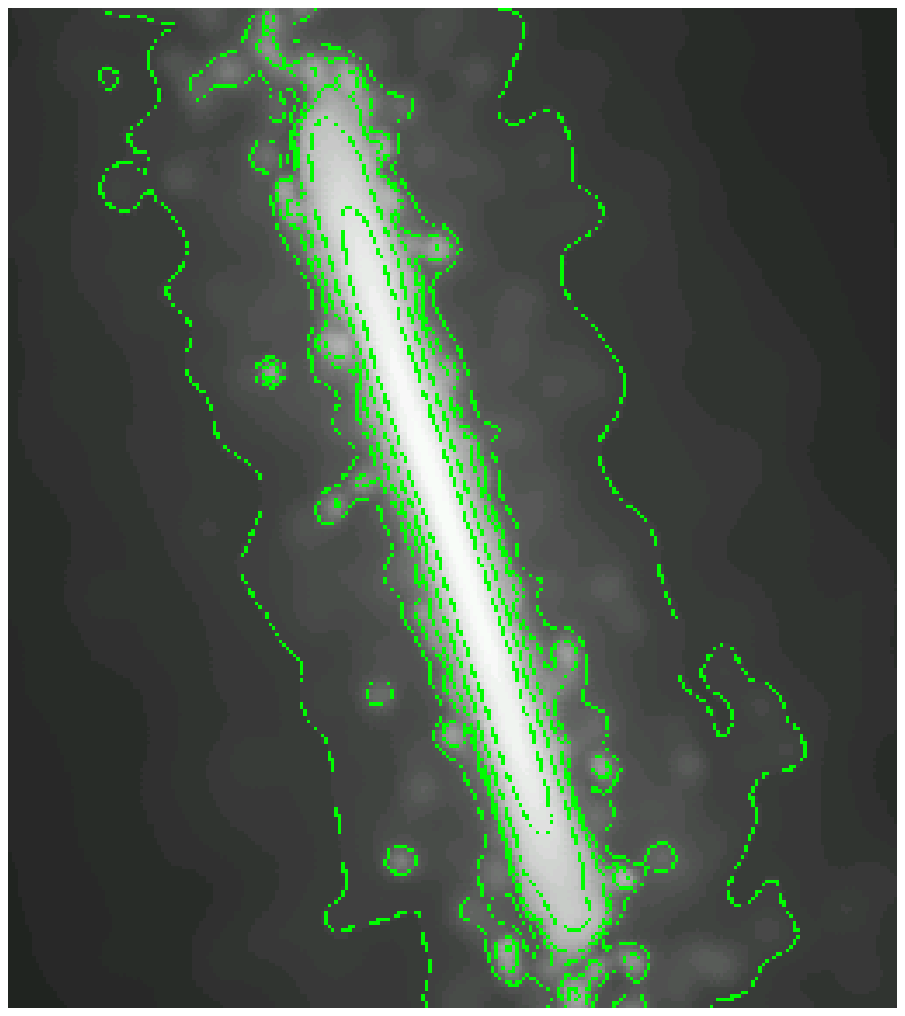} 
\caption{ 
The projected gas density of  
the standard M33 model after 1.5 Gyr shows a clumpy/irregular outer contours 
similar to that observed in NGC 891. The outer contour shows a  gas 
surface density of $\sim 10^{20}$ atoms/cm$^{-2}$. 
\label{cap:contours}} 
\end{figure}

\begin{figure*} 
\includegraphics[%
  scale=0.724]{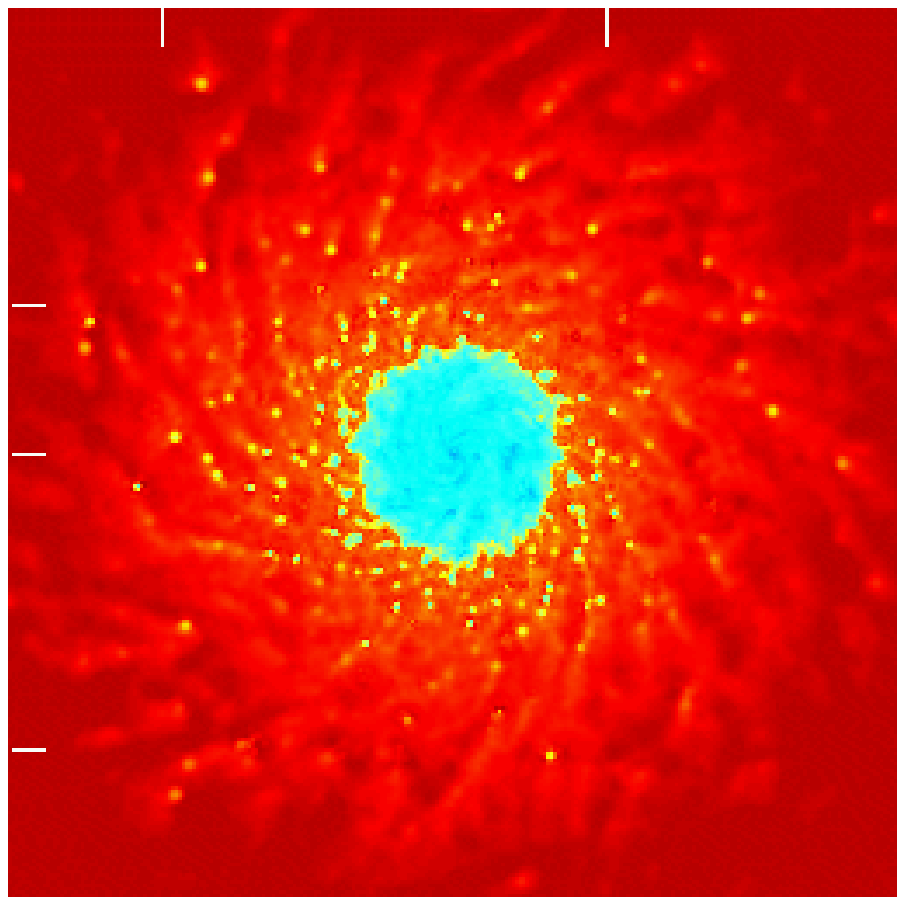} \includegraphics[%
  scale=0.75358]{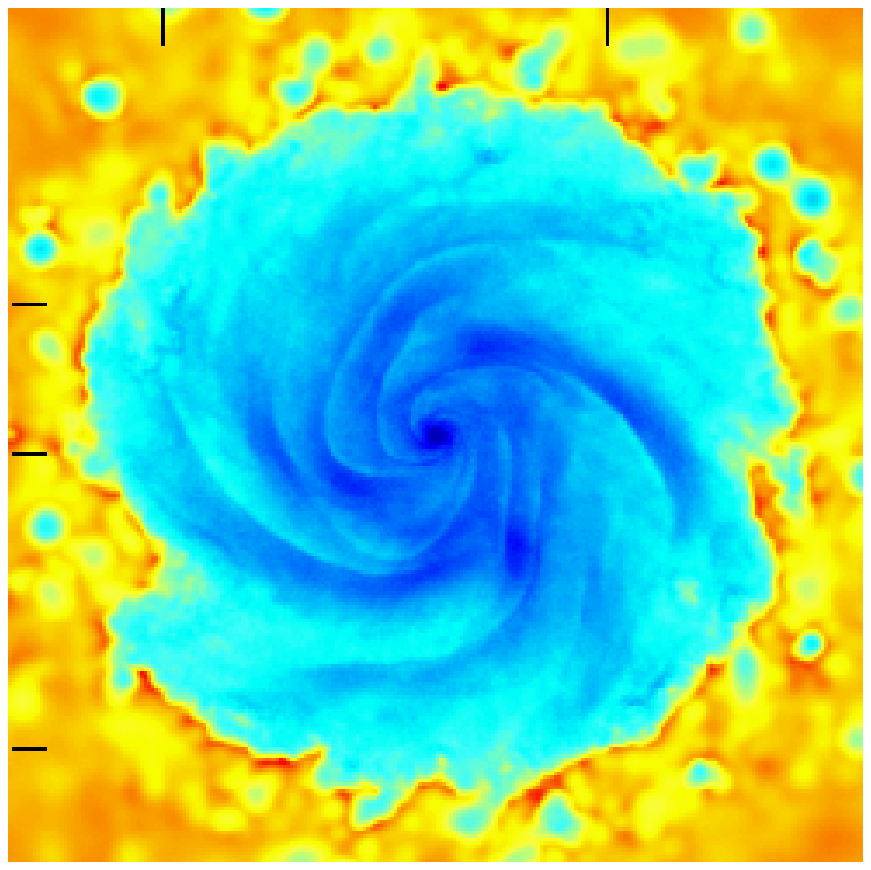}\includegraphics[%
  scale=0.39]{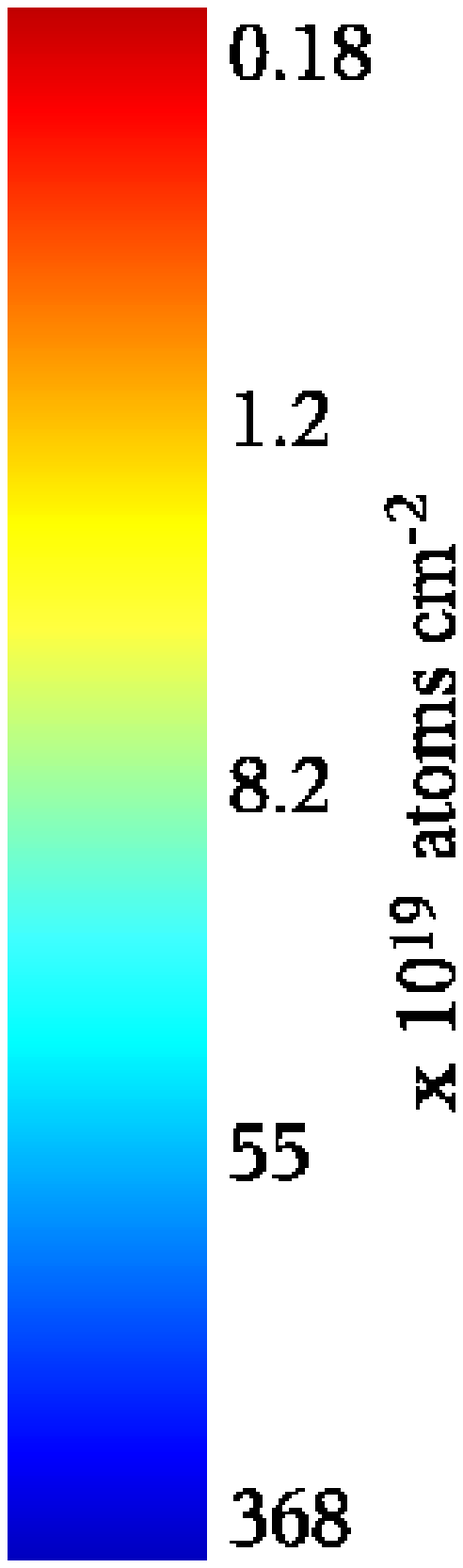}

\caption{ 
Density maps of gas in a slice through the centre of the M33 gas disk in boxes of length 40 kpc  
(left panel: refined\_8 resolution simulation after $0.5$ Gyr) 
and 20 kpc (right panel: standard resolution after $2.1$ Gyr ). The high-order
spiral pattern closely resemble that of Sc/Sd galaxies.  
\label{cap:blobs}} 
\end{figure*}

In the M33 model we can achieve a higher resolution since the lower density 
cold component requires less computational work. In the standard M33 model the 
force resolution is $100$ pc or $\sim0.06\%$ 
of the virial radius and the SPH particle mass is $2\times 10^4$ M$_{\odot}$. 
These simulations produce a large population of warm 
($T \sim 10^4-10^5$ K) high-density, pressure-confined clouds 
that are forming within the hot ($T \sim 10^6$ K) gaseous halo  
(see Figures \ref{cap:contours} and  \ref{cap:blobs}). These clouds eventually cool down 
to the temperature floor in our cooling function and would contain 
substantial fractions of neutral hydrogen. We will discuss the behaviour below the temperature floor in section \ref{belowten}. These clouds are in pressure 
equilibrium with the halo gas - their gravitational binding energy is  
typically $50$ 
times smaller than the thermal energy, hence the clouds are not 
gravitationally bound. The clouds have radii of $0.1-0.6$ kpc and masses 
from $10^5$ to a few $10^6 $ M$_{\odot}.$  As expected,  a lower number of clouds is found for a higher number of particles in the SPH smoothing kernel, but at the standard resolution we still see a cloud smoothing over $64$ neighbours (Table 1).
The parameters of the different simulations and the properties of the clouds 
are shown in Table 1.

\begin{figure*} 
\includegraphics[%
  scale=0.3548]{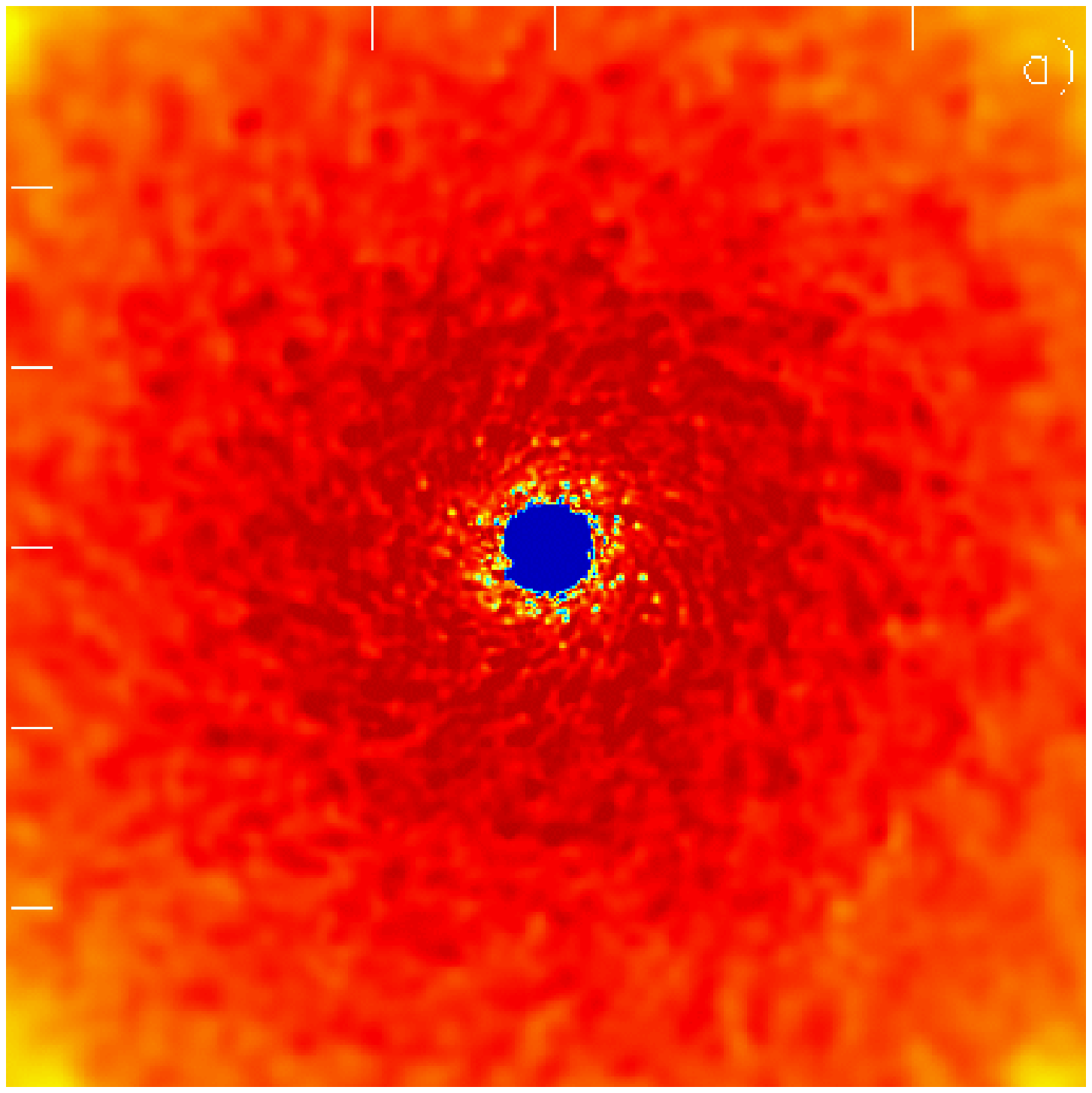}\includegraphics[%
  scale=0.605]{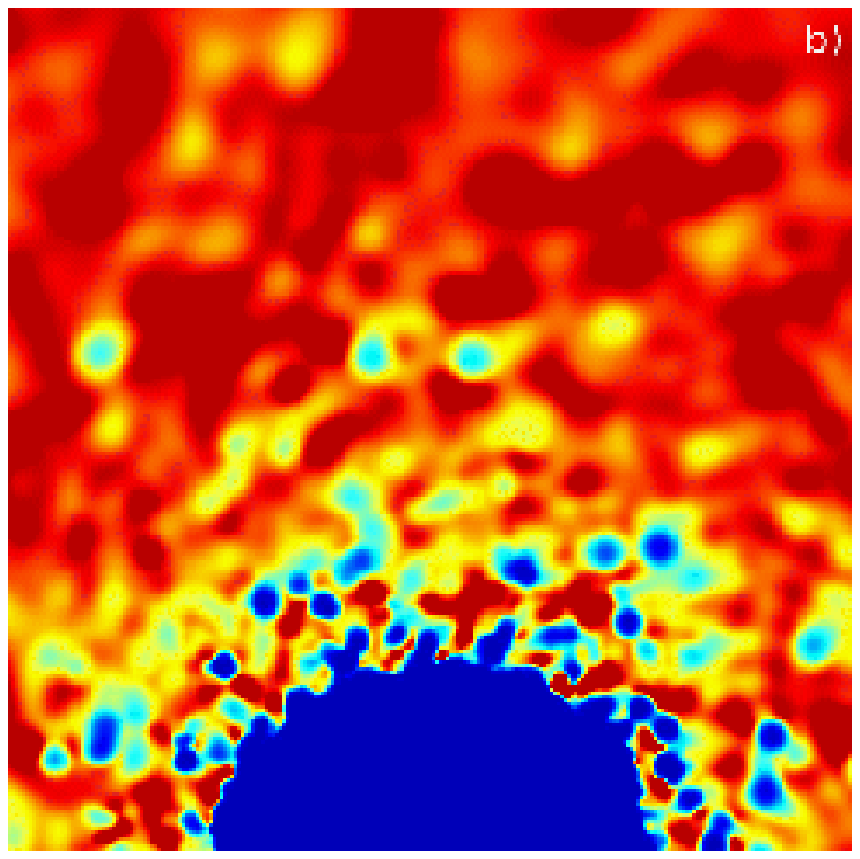}\includegraphics[%
  scale=0.394]{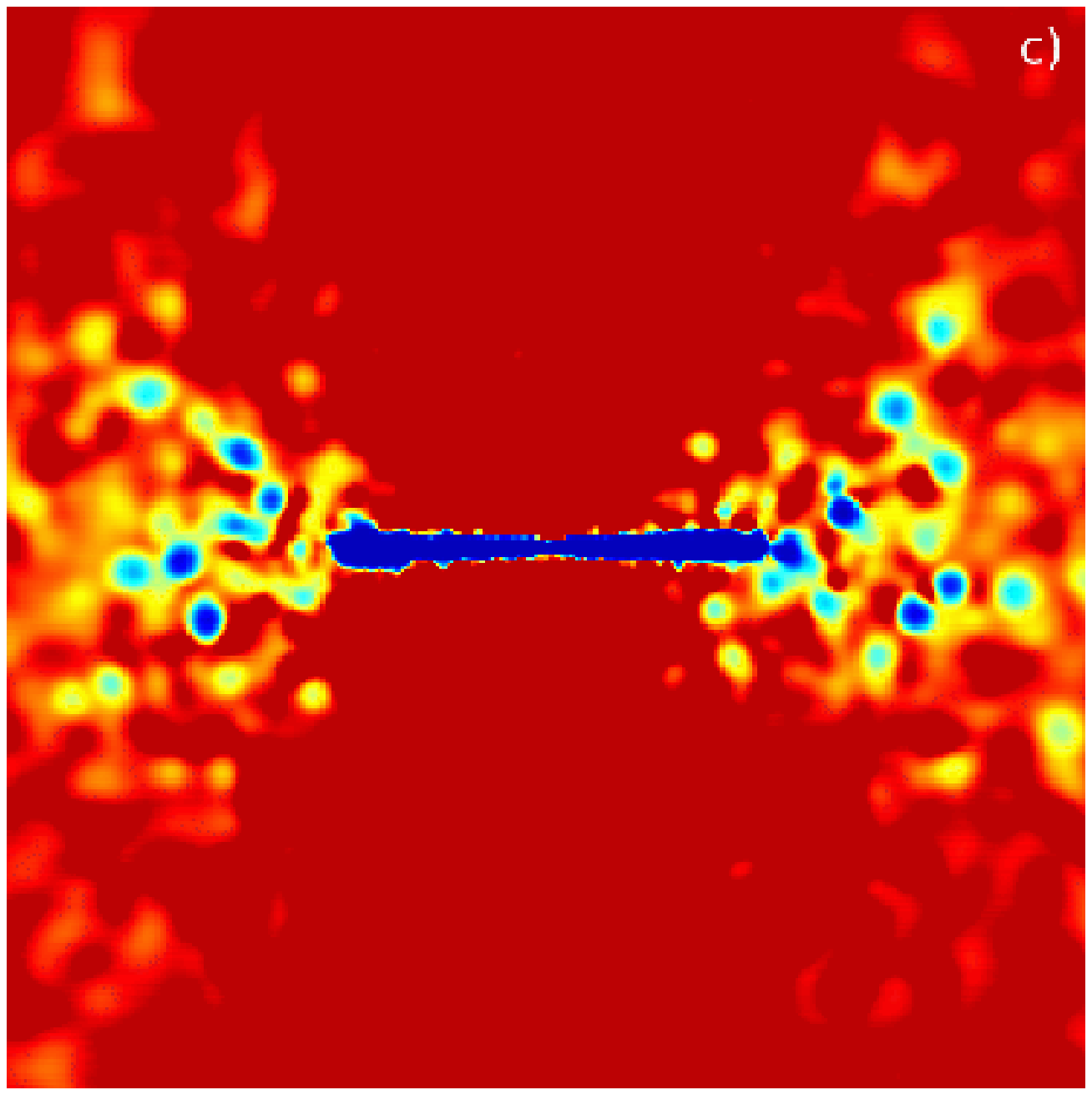}\includegraphics[%
  scale=0.310]{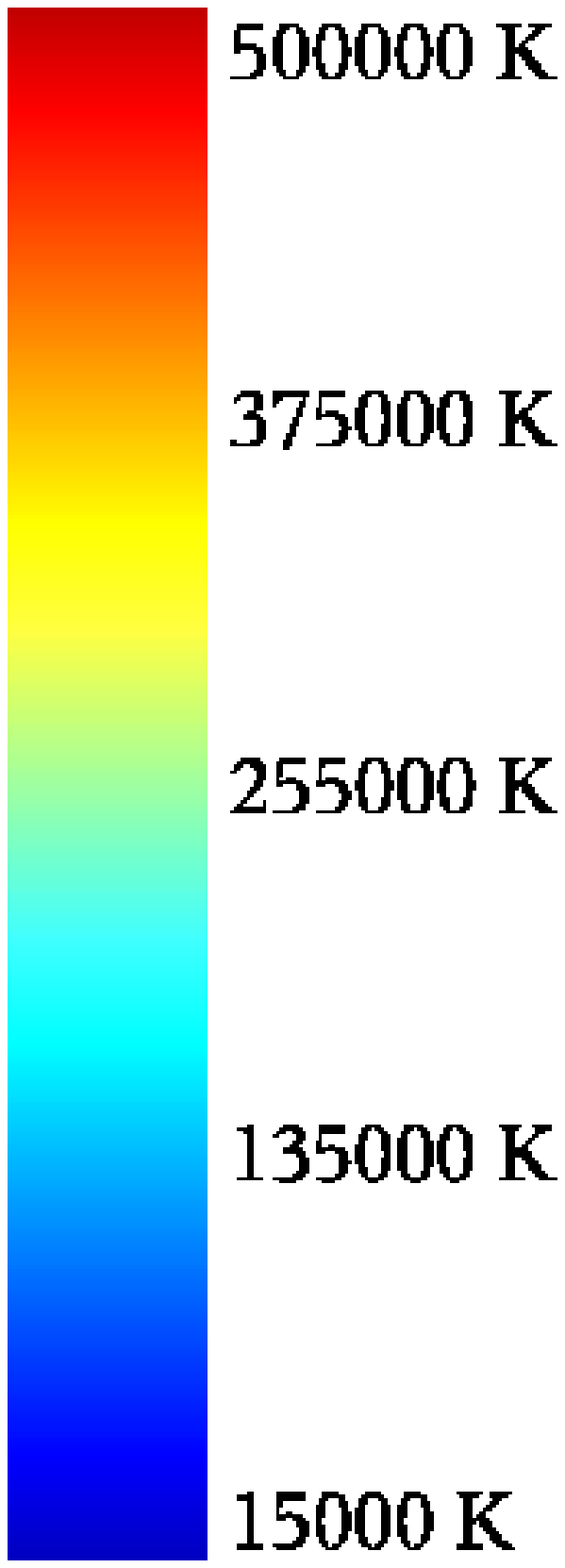}

\caption{Temperature maps of the gas in slices through the centre of the  
 standard M33 simulation after $2.1$ Gyr. Panel a) and b) show 200 kpc and 32 kpc 
regions respectively. Panel c) shows a 40 kpc slice perpendicular to the disk plane. 
\label{cap:tempblobs}} 
\end{figure*}

The thermal instability starts because of small density fluctuations  
present in the initial SPH particle distribution. 
These fluctuations rapidly grow, increasing in size and density until 
a pressure confined cold cloud forms. 
In reality these fluctuations might be driven by external 
perturbations, for example infall of cold gas and substructure or turbulence 
induced by supernova winds from disk stars or a galactic fountain. 
The temperature fluctuations are clearly visible throughout the gaseous halo in  
the left plot of Fig. \ref{cap:tempblobs}. These  
cold pressure confined clouds are seen only within about $10$ to $20$  kpc from the disk.  
This may be an SPH resolution effect - further out in the 
halo the gas density is lower and the SPH smoothing length is much larger. Thus density and 
temperature fluctuations are suppressed by the smoothing.
 
One of the conditions for cloud formation is that the sound-crossing time, 
$\tau_{\lambda}\simeq\lambda_{i}/v_{s},$ across a perturbation of 
wavelength $\lambda_{i},$ should be less than the characteristic 
cooling time (MB04). If this condition is not satisfied the perturbation 
is erased because the local cooling time is too close to the mean cooling time
and the whole region becomes   
isothermal. In our simulations we can 
resolve perturbations in the gas to $\sim2h,$ where $h$ 
is the smoothing length in the SPH code. We tested this condition 
in regions around the disk ($20$ kpc from the centre) where no clouds 
formed and found that $\tau_{\lambda}>\tau_{cool}=\frac{E}{\left|\dot{E}\right 
|}=\frac{T}{\left|\dot{T}\right|}$ (here we took $\lambda_{i} \sim 2h$) 
even at quite low temperatures ($T\approx30,000$ K). For regions where clouds 
form, the condition $\tau_{\lambda}<\tau_{cool}$ 
is fulfilled where the values of the smoothing length  
$h\approx0.65$ kpc.  
 
\begin{figure*} 
\includegraphics[%
  scale=0.755]{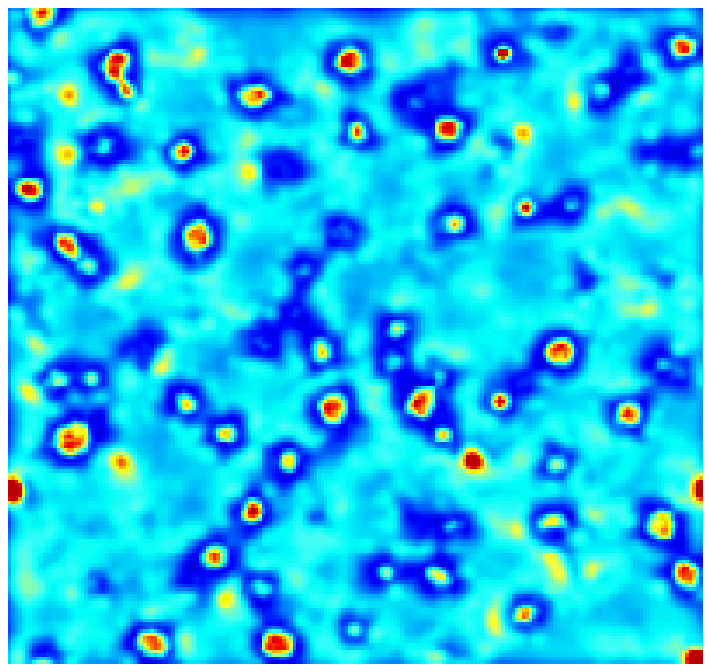} 
\includegraphics[%
  scale=0.3]{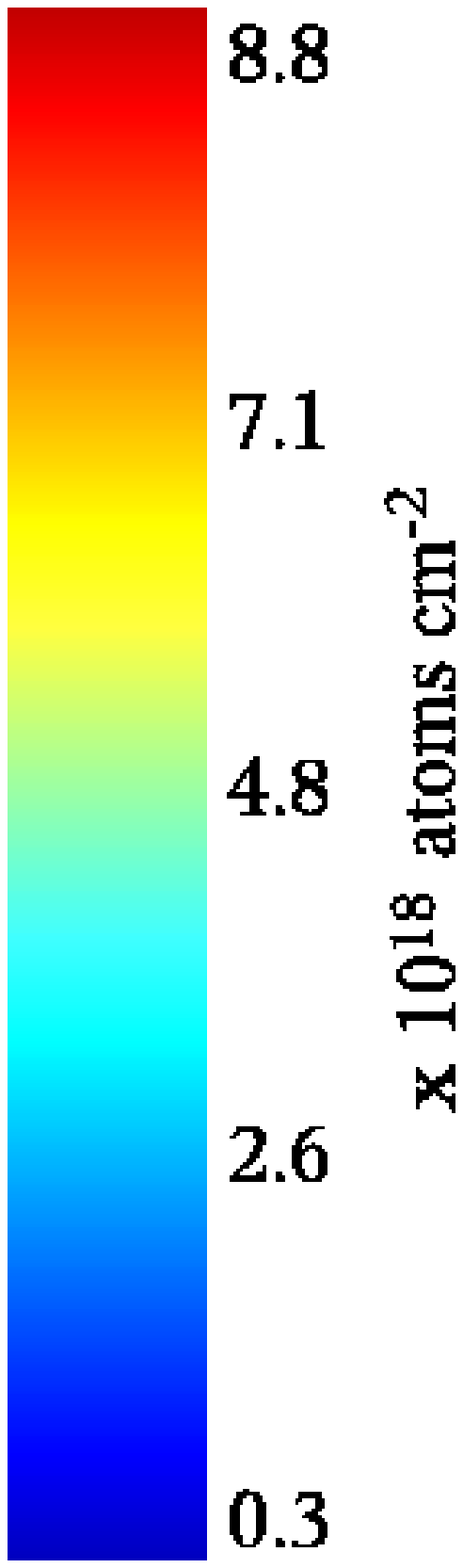}
\includegraphics[%
  scale=0.261]{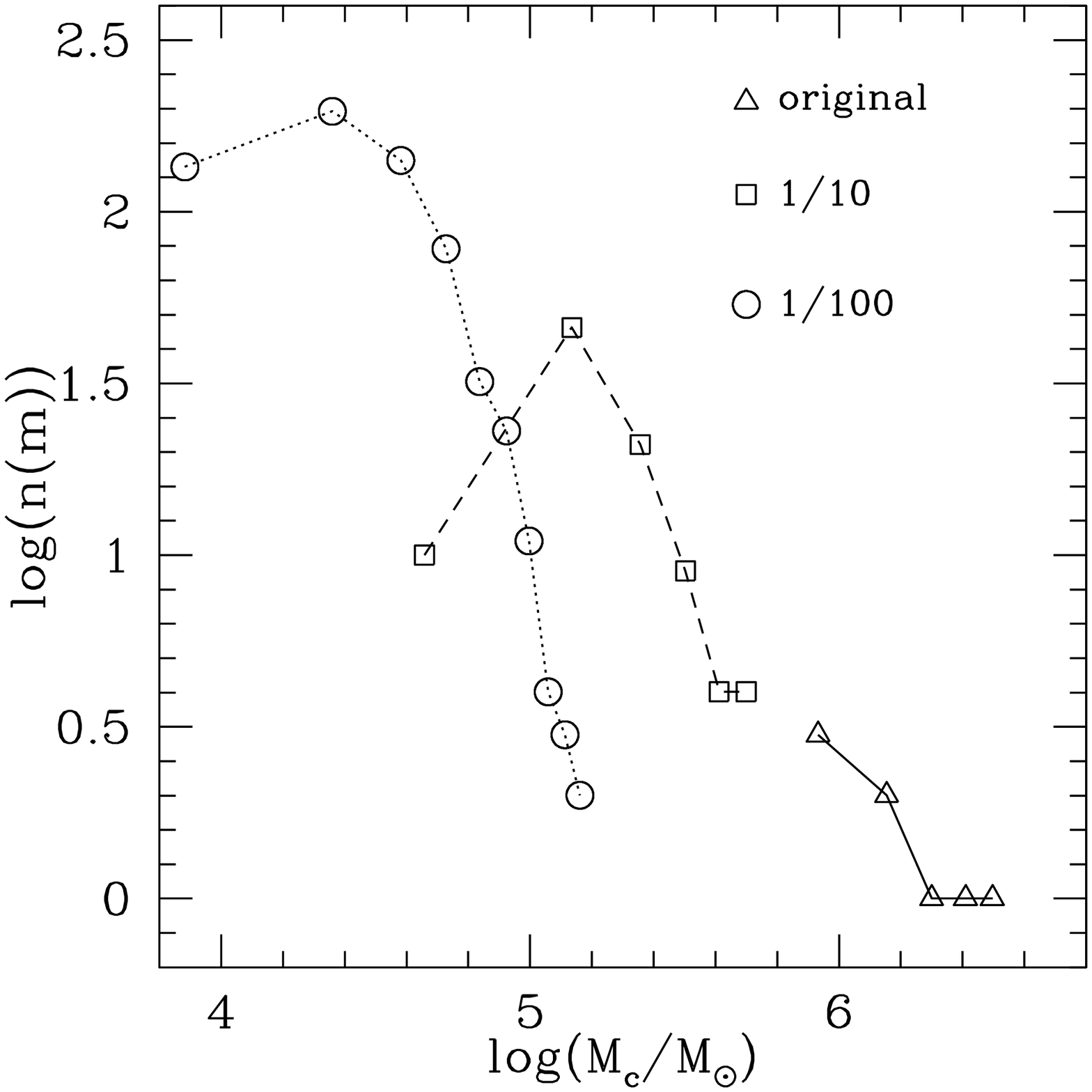} 
\includegraphics[%
  scale=0.261]{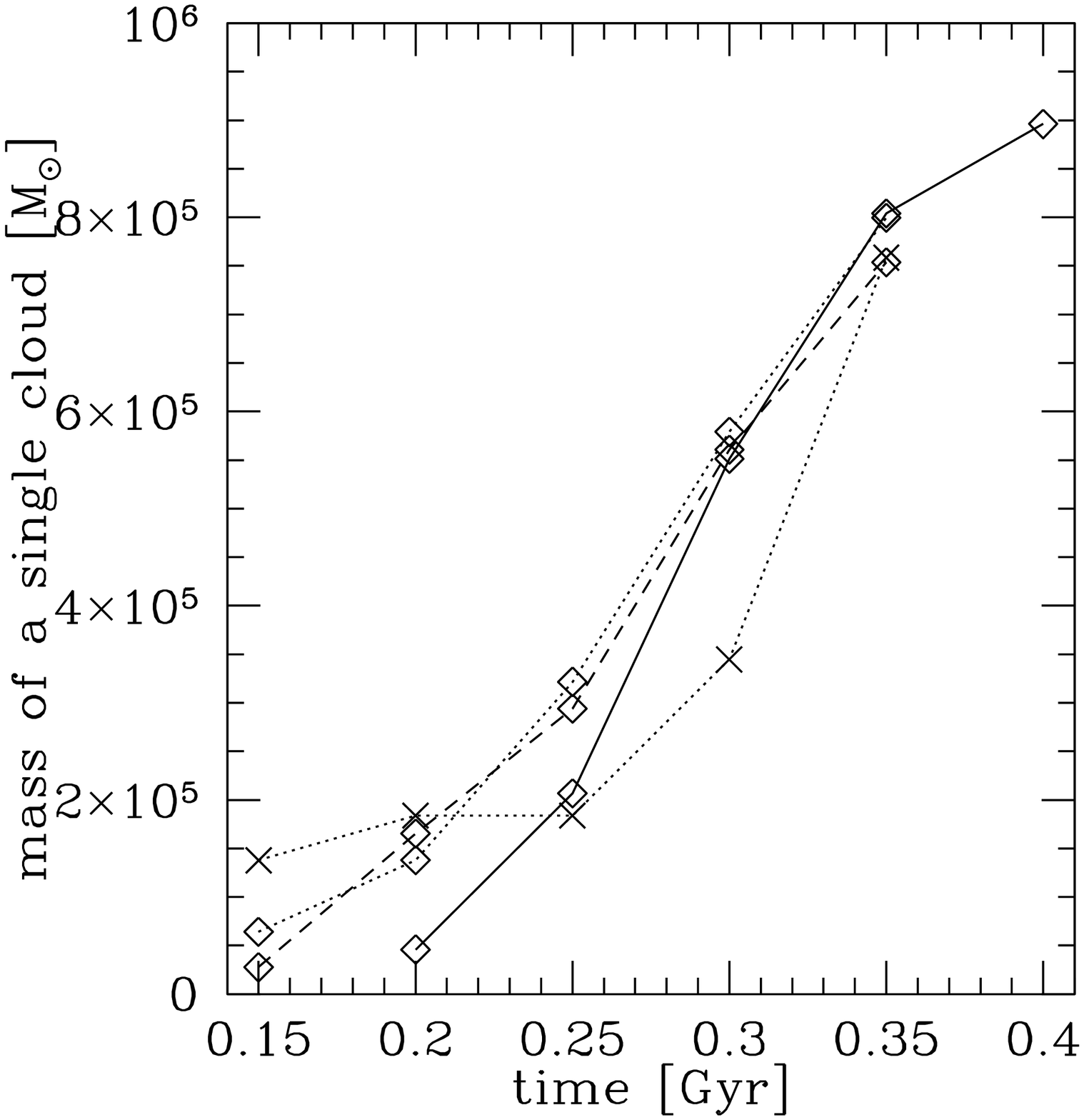} 
\caption{\label{cubes} The left panel shows a colour density map of a slice through the 
 centre of the highest resolution cube simulation after $0.4$ Gyr, where red indicates denser  
regions. The central panel shows the mass distribution of the clouds after $0.4$ Gyr The right panel shows the mass growth rate of a single cloud: The cube simulations are plotted  
with open symbols where the solid lines show the original resolution, dashed line intermediate scaled by a factor 4 and  dotted line is the highest resolution scaled by a factor of 40. 
A cloud from the standard M33 simulation is marked with crosses.  
In the latter case growth via merging begins after $0.25$ Gyr.} 
\end{figure*}

In order to assess the growth rates and resolution dependence of the instability
we performed simple tests using SPH particles distributed randomly  
in a 10 kpc periodic cube with initial temperature ($T=450,000$ K) and  
density ($n_H=0.01$ $cm^{-3}$) comparable to those from the central region of the halo models. 
We ran three simulations with particle masses 1, 1/10th and 1/100th times 
those in the full halo simulations. We found cold gas clouds forming above the 32 particle SPH smoothing kernel 
which grow linearly with time and mass (see Figure \ref{cubes}).  
The higher resolution simulations produce the smallest clouds since they have  
lower average Poisson fluctuations. After $0.8$ Gyr the whole box thermalises as the medium surrounding 
the clouds 
has had time to cool down to roughly the same temperature of the clouds. At 
this point the clouds dissolve into the outer medium since they are not  
pressure confined any more and the instability saturates.

\begin{figure} 
\includegraphics[%
  scale=0.4]{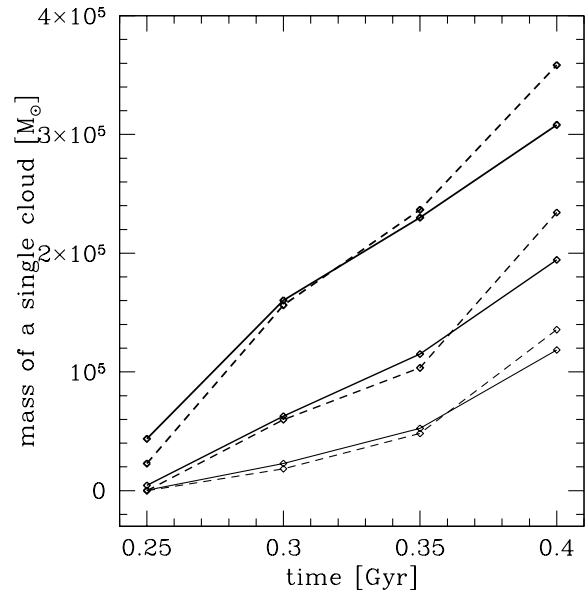} 

\caption{\label{ovdcubes} The mass growth rate of a single cloud in the box simulation with spherical overdensity is shown: The highest and intermediate resolution are plotted with solid and dashed lines, respectively. The thicker the lines the bigger was the initial overdensity ($10\%$,    $20\%$ and  $40\%$ ). The mass growth was close to linear and the results for the different resolutions nearly converge.} 
\end{figure}

We can compare the growth rates to those in Burkert \& Lin (2000), which derived a condition for exponential versus linear  
growth in a static, thermally unstable gas layer: the ratio of the initial cooling to sound-crossing time-scale $K_{j}$ has to be  
larger than the critical value $K_{crit},$ where 
\begin{equation} 
K_{j}=\tau_{cool}(0)k_{j}\sqrt{R_{g}T_{0}(0)/\mu},
\end{equation}
\begin{equation}
K_{crit}=\left(\frac{\rho_{a}}{\rho_{0}}\right)^{(2\beta-3)\Gamma/(4-2\beta)},  \end{equation} 
and where  $\tau_{cool}(0),k_j,R_g,\mu,T_0(0),\Gamma,$ and $\beta$ are the characteristic cooling time scale at time $0$, wavenumber, gas constant, initial temperature, mean molecular weight, adiabatic index, and the exponent of the power law of the cooling rate, respectively. $\rho_{a}$ is the initial perturbation amplitude and $\rho_{0}$ the initial mean density. The paper by  Katz et al. (1996) shows the cooling rate of a primordial composition gas (as is used in GASOLINE), for $3\times10^4<T<10^6$  a power law with $\beta=-1$ gives a rough approximation; with $\Gamma=5/3$ we find $K_{crit}\approx24.5$ for simulation with the  $10\%$ overdensity. For $K_{j}$ we get $\approx 46.8$ using $\tau_{cool}(0)=0.04$ Gyr, $T_0(0)=450,000$ K and taking two times the mean smoothing length as the smallest wavelength which is resolved ($\lambda_{min}=0.16$). Therefore the calculated $K_{j}$ is not a mean $K_{j}$ but the maximal one, it's value is roughly  two times bigger than the critical value, however the mass (and density) growth remain linear over several cooling times (Figure \ref{ovdcubes}). An additional reason for this may be because around $10^5$ K the real cooling curve decreases over several ten thousands of Kelvin and would therefore damp an instability.

The thermal instability generated in the cubes is purely of numerical origin since it is seeded by the initial Poisson fluctuations. A higher
resolution setup simply starts with lower amplitude of the initial fluctuations, hence the different mass of the transient clouds formed.
The following step is thus to impose a perturbation and study how the system responds when we increase the resolution. This way
we can assess robustly the dependence on resolution since the initial perturbation is independent on resolution. We used the cubes  with particle masses 1/10th and 1/100th times those in the full halo simulations.  To wash out initial Poisson noise the cubes were evolved for $5$ Gyr using an isothermal EOS (with the same temperatures and mean density as before). Then we placed  a spherical overdensity at an arbitrary location within the box. We run three different models where the overdensity was on the $10\%$,    $20\%$ and  $40\%$ level, respectively, at both resolutions. We find that the mass of the clouds in this case, just before the whole box thermalises, differs roughly by the same 
factor as the initial overdensity, the growth rate is linear in time as in the previous cube tests, and the results in the two resolutions nearly 
converge (see Figure \ref{ovdcubes}). In passing we note that the results at $0.4$ Gyr probably overestimate slightly the cooled mass due to the 
overdensity, because in the next time-step the whole box has already thermalised. The fact that we found nearly convergent results has important
implications for the collapse simulations that we discuss below. In fact the intermediate resolution test has particles masses comparable 
to the ``refined\_8'' M33 simulation, which should thus have enough resolution to properly resolve the thermal instability for an initial
amplitude of the fluctuations $10\%$ or higher.

So far we had built-in or imposed perturbations as seeds for the thermal instability. We can ask if there is a reasonable way of producing
perturbations of order 10\% or larger, namely of the same magnitude of those existing in the initial conditions of the full simulations.
One way these perturbations can be produced is by dark matter and baryonic substructure in galactic halos.
We carried out an additional test run using the hires periodic cube. We added
a massive particle with characteristic size (i.e. gravitational  softening) and mass comparable to the Magellanic Clouds. The particle
moves through the box at a speed of $\sim 200$ km/s as expected for a dwarf galaxy satellite orbiting in the Milky Way halo.  
Starting with an isothermal EOS  we saw that the particle rapidly triggered the formation of a trailing overdensity well above $20\%$ of  the background density just behind. Evolved with cooling this 
overdensity grew considerably faster than the fluctuations due to numerical noise, as expected from its larger amplitude.   
Clearly we have chosen a favourable case, since most of the galactic satellites are lighter than the LMC and would drive weaker
perturbations.  Triggering of gas density fluctuations by substructure in a CDM halo should indeed be dominated by the many dark satellites at 
least 100 times lighter than the LMC. Individually, these would drive perturbations that are too small to grow significantly on a dynamical
time. However, perturbations presumably would not be isolated but would interact, possibly amplifying more rapidly than expected from the
linear growth. This is already suggested by the fact that clouds in the full simulations grow faster than in the cube tests due to
merging.  The presence of a clumpy medium near the disk, which we 
invoked to explain the observations of Fraternali et al. in previous sections, would suggest 
that there must be a source of the perturbations acting near the disk, fairly isotropic and 
active even at the present epoch. Whilst satellite accretion of fly byes may have been frequent enough early
in the galaxy formation process to be a major source of perturbations near
the disk, other mechanisms are probably responsible for triggering the thermal 
instability today. Galactic fountains driving turbulence 
in the inner part of the hot gaseous corona might be one possibility. We will address the question on the physical origin of those perturbations in an upcoming paper.

There are important differences between all the cube tests performed and the full
simulation. One that we just mentioned is that in the full simulations clouds can grow faster due to mergers with
other nearby clouds (see Figure \ref{cubes}).  Another one is that the perturbations are not growing out of a static gaseous background 
as in the cubes but within infalling hot gas. The static case was recently investigated in Baek et al. (2005) where a perturbation at the 
 $20\%$ level was used to produce clouds with masses of about $10^5 $ M$_{\odot}$  
using a grid code. 
In a similar spirit, Fall \& Rees (1985) showed that an initial  
density perturbation of the order $10\%$ of the average density is needed  
to produce clouds with masses comparable to those of globular clusters in
an infalling medium. This is consistent with the fact that in the full simulations the initial perturbations are  
larger than $10\%$, so that one expects the thermal instability to be 
able to produce fairly massive clouds.
 
\begin{figure} 
\includegraphics[%
scale=0.4]{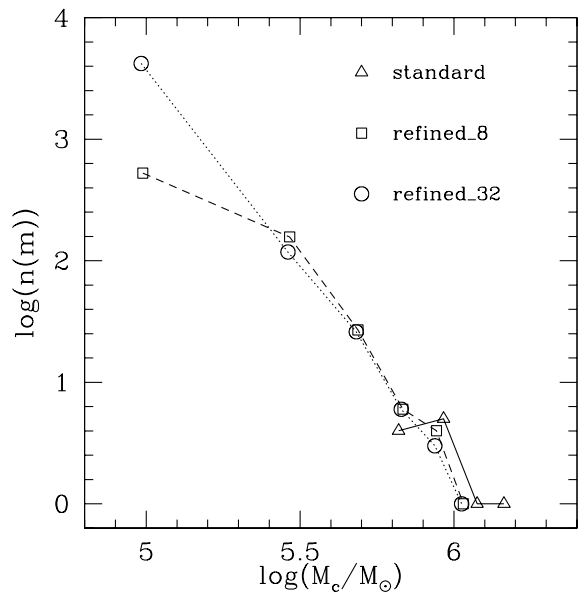} 
\caption{ The 
mass distribution of clouds in the M33 simulation after $0.5$ Gyr,  
where the  dotted, dashed and solid curves show the refined\_32, the refined\_8 and standard 
resolution simulations (see text and Table 1), respectively. \label{UMGM}} 
\end{figure}

 In order to test the effects of resolution directly in the full simulations we have run two simulations with particle
splitting. First we carried out the 
refined\_8 M33 simulation. Particles  within the refined sphere of 30 kpc reach the disk in about $1.5$ Gyr. By running the 
simulation only up to 
$1$ Gyr we make sure that there is no contamination of heavier particles in 
the disk, which can lead to spurious results (Kitsionas 2000).  
The splitting scheme should  
preserve the amplitude of perturbations at large scales that were present in   
the original simulation, while noise at previously unresolved scales 
should trigger the formation of smaller clouds.  After $0.5$ Gyr (see Figure \ref{cap:blobs} and  
Table \ref{cap:tab mass 2})  
there are about $700$ resolved clouds with mean mass  $1.8 \times 10^{5}$ M$_{\odot}$,   up to $19$ kpc from the centre. The most massive cloud has 
a mass of $1.1\times 10^{6}$ M$_{\odot}$, which is comparable to the most massive  
cloud ($1.6 \times10^{6}$ M$_{\odot}$) in the standard M33 simulation 
{\it at the same time}.  
We then performed an even higher resolution run by splitting the initial particle distribution by an additional factor of 4.
In the refined\_32 there were more than $4300$ clouds having more than $32$ particles due 
to the smaller perturbations. The large-scale perturbations were followed similarly in all these three simulations, resulting in a 
comparable mass distribution of the clouds at the high mass tail, see Figure \ref{UMGM}. 
The satisfactory convergence at the scales resolved by all simulations is not a trivial result; it shows that the amplification of the 
perturbations is numerically robust even in complex simulations such as these in which the gas density does not change simply because the local 
cooling rate changes but also as a result of compressions and shocks driven by the gravitational collapse. In other words only the
initial amplitude of the fluctuations is not self-consistently determined.

\begin{table*} 
\begin{tabular}{|c|c|c|c|c|c|c|c|} 
\hline  
\# {\footnotesize dark/gas particles}&
$m_{g}$ [M$_{\odot}]$ &
{\small $T_{cut-off}$ {[}Kelvin{]}}& 
Time/Gyr& 
$h_{min}$& 
$N_{SPH}$& 
number of clouds& 
total mass in clouds [M$_{\odot}]$\tabularnewline 
\hline 
\hline  
{\small $2.5\times10^{5}$/ $1\times10^{5}$}& 
$4.6\times10^{5}$&
$30000$& 
$1.5 $& 
-& 
$32$& 
0& 
0\tabularnewline 
\hline  
$1.1\times10^{6}/5\times10^{5}$& 
$9.2\times10^{4}$&
$30000$& 
$1.5 $& 
-& 
$32$& 
2& 
$9.01\times10^{6}$\tabularnewline 
\hline  
$2.2\times10^{6}/2\times10^{6}$& 
$2.3\times10^{4}$&
$30000$& 
$1.5 $& 
-& 
$32$& 
13& 
$1.23\times10^{7}$\tabularnewline 
\hline  
$2.2\times10^{6}/2\times10^{6}$& 
$2.3\times10^{4}$&
$30000$& 
$1.5 $& 
-& 
$64$& 
1& 
$2.32\times10^{6}$ \tabularnewline 
\hline  
$2.2\times10^{6}/2\times10^{6}$&
$2.3\times10^{4}$&
$15000$& 
$1.5 $& 
-& 
$32$& 
13& 
$1.52\times10^{7}$\tabularnewline 
\hline  
$2.2\times10^{6}/3.9\times10^{6}$&
$2.9\times10^{3}$& 
$15000$& 
$0.5 $& 
-& 
$32$& 
410& 
$7.4\times10^{7}$\tabularnewline 
\hline
$2.2\times10^{6}/1.0\times10^{7}$&
$7.3\times10^{2}$& 
$15000$& 
$0.5 $& 
-& 
$32$& 
3662& 
$2.3\times10^{8}$\tabularnewline 
\hline  
$2.2\times10^{6}/2\times10^{6}$&
$2.3\times10^{4}$& 
$15000$& 
$0.65 $& 
-& 
$32$& 
7& 
$8.46\times10^{6}$\tabularnewline 
\hline  
$2.2\times10^{6}/2\times10^{6}$& 
$2.3\times10^{4}$&
$15000$& 
$0.65 $& 
$20$ pc& 
$32$& 
11& 
$1.03\times10^{7}$\tabularnewline 
\hline 
$2.2\times10^{6}/2\times10^{6}$&
$2.3\times10^{4}$& 
$30000$& 
$1 $& 
- & 
$32$& 
36& 
$4.2\times10^{7}$\tabularnewline 
\hline 
\end{tabular}\vspace{0.375cm}

\caption{Simulation properties of the M33 models and cloud statistics for clouds identified with a friends of friends algorithm are shown. The minimum number of particles for objects to be identified with the FOF was set to be $32$, in addition we used also $64$ particles for the run with $N_{SPH}=64$, which gave the same result. The standard,  refined\_8 and refined\_32 simulations are shown, respectively, in the fifth, sixth and seventh row. Rows eight and nine show two equal simulations differing  
only in the use of a minimal smoothing length. The final row shows the simulation with the enhanced cooling rate. The mass quoted in the second column is referring to the mass of one gas particle initially near the centre.\label{cap:tab mass 2}} 
 
\end{table*}

\subsection{Evolution and dissolution of the clouds} 
 
In the standard M33 run, a typical compact cloud has a mass of $\sim1.1\times10^{6}$ M$_{\odot}$ and a temperature equal to the cut-off temperature in the cooling function  
($T_{cloud}=15,000$ K), radius $R\sim0.2$ kpc and lifetime of $\sim0.1$ 
Gyr (from formation to entering the disk). We now investigate the 
mechanisms that could disrupt or modify the structure of cold clouds 
embedded in a hot medium. 
 
\subsubsection{Kelvin-Helmholtz Instability} 
 
As a cool, dense cloud moves through a hot, tenuous background, the 
interface between the two phases is subject to the growth of Kelvin-Helmholtz 
instabilities. At our resolution, SPH does not resolve such instabilities due to smoothing 
and the artificial viscosity which tends to blur any sharp interface between 
the inner and outer medium. In other words, the Reynolds number is always lower 
than the Reynolds number expected in turbulent flows where the Kelvin-Helmholtz 
instability can develop (see Mayer et al. 2006).  
For the case in which gravity is unimportant Murray et 
al. (1993) derive a characteristic growth time for the instability of 
\begin{equation}\tau_{g}\approx\frac{R_{cloud}(\rho_{cloud}/\rho_{bg})^{0.5}}{U},\end{equation} 
where $U$ is the relative velocity and $\rho_{bg}$ is the density 
of the background medium. The clouds are therefore expected to break 
up on time-scales comparable to $\tau_{g}$. In their numerical experiments 
they actually found that the mass loss was still quite small over timescales 
twice as long as $\tau_{g}.$ 
In our simulations the density contrast is $\approx100$, the relative velocity $U$ is $\approx10$ km/s and therefore we end up with
$\tau_{g}\approx0.2$ Gyr using $0.2$ kpc for the radius of the cloud. Since such timescale is longer than the typical
lifetime of the clouds we expect the effect of the Kelvin-Helmholtz 
instability to be negligible. 
 
The radial infall velocities is of the order $10$ km/s 
which is insufficient to perturb the cloud structure. As the halo gas density becomes 
lower we might expect clouds to infall faster. Observations of the radial motion 
of clouds could therefore be used to constrain the ambient halo density. 
 
\subsubsection{Conduction} 
In principle, conduction can prevent the growth of small clouds.  MB04 suggest  
random magnetic fields reduce conduction by at least an order 
of magnitude (the reduction could be much stronger if the field is uniform or 
tangled) setting a minimum cloud mass of about $10^{5} $ M$_{\odot}$ for a galaxy 
with mass comparable to that in our M33 model. 
For the standard simulation presented here, the conduction limit thus lies below 
the resolution limit imposed by the SPH smoothing volume (and slightly above 
the resolution limit for the simulation with splitting in 32 particles. The number of small clouds is therefore expected to be lowered due to conduction). At very high  resolution, conduction would provide a natural cut-off scale and allow detailed numerical convergence.

\subsubsection{Cooling below $10^4$ K?} \label{belowten} 
 
Although we used a lower cut-off in the cooling function to prevent 
fragmentation it is interesting to investigate the fate of the clouds 
below that temperature. In order to do that we used a ``standard''  
cooling function (e.g. Dalgarno \& McCray 
1972) to derive an estimate for the cooling timescale 
of the clouds below $10^4$ K.
We adopted the parameterisation of Gerritsen \& Icke (1997) for the temperature 
range of $\log T<6.2$:  
 
\begin{equation} 
\Lambda=10^{-21}n_{H}^{2}[10^{-0.1-1.88(5.23-\log T)^{4}}+10^{-a-b(4-\log T)^{2}}]\end{equation} 
in ergs cm$^{-3}$ s$^{-1},$ where $a$ and $b$  
depend on the ionisation parameter $x=\frac{n_{e}}{n_{H}}.$ The heating 
term for photoelectric heating of small grains and PAHs (as the largest 
contributor to the gas heating) is given as $\Gamma=10^{-24}\epsilon G_{0}n_{H}$ 
in ergs cm$^{-3}$ s$^{-1},$ where $\epsilon$ is the heating efficiency 
and $G_{0}$ the incident far-ultraviolet field (Bakes \& Tielens 
1994). The FUV field originates from star formation in the disk; since we 
are resolving only clouds very close to the disk we can simply assume that 
they will be embedded in the same bath of ionising photons as the rest 
of the galaxy. 
With \begin{equation} 
\frac{du}{dt}=\frac{\Gamma-\Lambda}{\rho}\end{equation} 
(Hernquist \& Katz 1989) one can give an estimate of the cooling time 
for such systems neglecting pressure and velocity terms. A typical 
cloud in our simulation can cool down during its lifetime to several 
tens of Kelvin for ionisation parameters $x\geq0.1$ in less than 
$1$ Myr, for $x\leq0.01$ an equilibrium is reached at several thousand 
Kelvin. The dynamical time-scale $t_{dyn}=\sqrt{\frac{3\pi}{16G\rho}}\approx70$ 
Myr for such a cloud is of order of its lifetime. Therefore the clouds 
could collapse further for $x\geq0.1$ ( a value which should be reasonable 
for the solar neighbourhood, see Cox 1990). 
On the other hand MB04 find that the 
extragalactic ionising background should prevent the cooling of the 
clouds below $10^{4}K$. Self-shielding effects will also be 
important to decide the ultimate fate of the clouds under the combined 
action of different heating and cooling mechanisms.  
 
Had we included metals in the simulations the cooling rate would have 
been higher. We crudely explored 
the effect of metals by simply shifting the entire cooling curve up.  
We found, as expected, that more efficient cooling leads  
to a larger number of clouds for a given 
resolution; doubling the energy loss due to cooling  
(this modified cooling function is somewhere  
in between primordial gas and gas with solar metalicity)  
increased the mass and the number in clouds by a factor  
of order $2.7$ whereas the mass per cloud and the  
spatial extent of the clouds remained comparable. 
 
\section{Conclusions} 
 
We studied the cooling flow of gas within equilibrium dark plus gaseous haloes. 
At high resolution the gas nearly conserves  angular momentum and rotates faster as it moves 
towards the disk. Within $10-20$ kpc from the disk the gas fragments into cold gas clouds 
pressure confined by the ambient hot halo gas. The clumpy appearance of the gas matches HI observations of outer disks. These cold clouds have similar properties as  
predicted in analytic models such as that of MB04. However, one difference  
with MB04 is that the hot gas component is never in approximate hydrostatic equilibrium once it starts cooling and
collapsing. This implies, for example, that the disk assembly will occur faster than predicted by a model
such as MB04. The mass of galaxies with a given spin should also be different, presumably higher in our case
since there is less pressure support against collapse. Unfortunately a direct comparison in terms of galaxy masses
is not possible at this stage since the refined runs that have enough resolution to properly follow the clumpy
disk formation have been run only for a very short timescale.
The background gas cools and falls 
in towards the disk as it loses pressure support. Its radial infall 
velocity is comparable to that of the cold clouds. This explains why we do not see significant differences in the mass growth of the disk when we compare simulations which yield a very different degree of clumpiness of the accretion flow such as the standard and refined\_32 simulations.
In our models the amount of hot gas which is accreted directly to the disk  is  substantial, however, even after $5$ Gyr, there is a residual hot gas halo, such that a large mass fraction of the baryons remain in the halo. 

The gas does not accrete onto the disk via a spherically symmetric flow. Rather it flows down a cylindrical fashion along the angular momentum axis. This is different 
from that assumed in semi-analytic models for disk formation. 
The rotational velocity of the gas decreases above the disk plane with a  
similar velocity gradient as observed in the measurements for NGC 891 by  
Fraternali et al. (2005). This may be the first evidence for disk formation 
via radiative cooling of hot gas combined with conservation of angular momentum. 

Our simulations show that the thermal instability, and thus the two-phase medium, requires a continuous heating source to be present in order to be maintained, and this is provided here by the gravitational collapse which continously brings in gas hotter than the
gas that has already cooled down. When this heating term is not present, such as in all the cube tests, the instability
is rapidly suppressed as the temperature becomes uniform due to the high cooling rate.

We believe that the clouds are stable against the various disruption mechanisms 
such as conduction and Kelvin-Helmholtz instabilities. The infall velocity  
of the gas is approximately $10$ km/s. This is much smaller than the pure free-fall 
velocity towards the disk, $\sim 70$ km/s, owing to the drag exerted by the 
diffuse hot corona. This velocity can thus be used to measure the 
ambient hot gas density above the disk plane. This radial infall velocity  
agrees well with that observed by Fraternali et al. (2002) of   
approximately $15$ km/s for NGC 2403, a galaxy similar to M33.

In the refined\_32 simulation the sky coverage factor of the clouds is approximately 25\%. This is quite
similar to the sky coverage factor of high column density High Velocity Clouds (HVCs) (Richter 2006).
The number  of  cold clouds in our (refined\_32) simulation is also similar to those estimated
by Putman et al. (2006). However, we caution against over-interpreting these intriguing results.
In our model several physical ingredients that might have an impact on cloud formation, such as gas metalicity,
and survival (such as conduction) are still missing, and the initial perturbations are still not determined
self-consistently in the mode. It is interesting that cosmological simulations are now beginning to resolve
similar structures (Governato et al. 2006, Sommer-Larsen 2006) and that the origin of the perturbations 
in some of these simulations seems to be dense cold gaseous streams accreted in the process of galaxy formation 
(Sommer-Larsen 2006). If cold accretion is responsible for the cloud formation it would have the benefit
of being a very general mechanism, although it would occur preferentially in the early stages of their formation
for systems as large as the Milky Way today (Kere{\v s} et al. 2005).

The growth rate of the cloud masses  
is approximately linear with time, therefore large initial  
fluctuations in temperature or density are required. It is likely that the Poisson  
fluctuations due to discreteness, which are at the 10\% level, are  
coincidentally close to the amplitude of  
fluctuations required to form $10^6$ M$_{\odot}$ clouds within a Gyr. 
In addition to perturbations of cosmological origin for the clouds to be connected with HVCs one needs mechanisms
that are efficient at any epoch, such as triggering by halo substructure. This includes not only the direct
perturbation of sub-haloes moving in the primary halo but also cold gas streams stripped from the baryonic
component of the satellites (Mayer et al. 2006).
Finally, in order to generate the HI clouds in the extra planar gas around disks one needs also local
sources of fluctuations  such as infalling cold gas or turbulence induced by gravitational 
perturbations or supernova winds from star formation in the disk.  
A lot remains to be done to study the role of these different triggering mechanisms which might be simultaneously 
at play and generate various populations of clouds.

\section*{Acknowledgements} 
We would like to thank Stelios Kazantzidis for providing a code to generate isolated dark matter haloes  
and Filippo Fraternali for the rotation curve data for NGC 891.
We acknowledge useful and stimulating discussions with Filippo Fraternali, Renzo Sancisi, Frank van den Bosch, Kenneth Sembach, Leo Blitz, Tom Quinn, Fabrizio Brighenti and  Fabio Governato. We are grateful to the anonymous referee for his remarks and stimulating comments that considerably improved the paper.   
The numerical simulations were performed on the zBox  
(http://krone.physik.unizh.ch/$\sim$stadel/zBox)  
supercomputer at the University of Z\"urich.

\label{lastpage} 
 
\end{document}